\documentclass[sigconf]{acmart}

\usepackage{times,epsfig}
\usepackage{caption}
\usepackage{graphicx}
\usepackage{amsmath,amsfonts,amssymb}
\usepackage{bm}
\usepackage{multirow}
\usepackage{rotating}
\usepackage{url}
\usepackage{subfigure}
\usepackage{array}
\newcolumntype{M}[1]{>{\centering\arraybackslash}m{#1}}
\usepackage{amsmath}
\usepackage{color}
\usepackage{refcount}
\usepackage{comment}

\renewcommand\footnotetextcopyrightpermission[1]{} 
\setcopyright{none}

\begin{document}
\title{Styling with Attention to Details}

\author{Ayushi Dalmia, Sachindra Joshi, Raghavendra Singh, Vikas Raykar}
\affiliation{%
 \vspace{1.25em}\institution{IBM Research}
}
\affiliation{%
  \institution{\{adalmi08,\hspace{0.2em}jsachind,\hspace{0.2em}raghavsi,\hspace{0.2em}viraykar\}@in.ibm.research.com}
}

\begin{abstract}
Fashion as characterized by its nature, is driven by style. In this paper, we propose a method that takes into account the style information to complete a given set of selected fashion items with a \textit{complementary} fashion item. Complementary items are those items that can be worn along with the selected items according to the style. Addressing this problem facilitates in automatically generating stylish fashion ensembles leading to a richer shopping experience for users. 

Recently, there has been a surge of online social websites where fashion enthusiasts post the outfit of the day and other users can like and comment on them. These posts contain a gold-mine of information about style. In this paper, we exploit these posts to train a deep neural network which captures style in an automated manner. We pose the problem of predicting complementary fashion items as a sequence to sequence problem where the input is the selected set of fashion items and the output is a complementary fashion item based on the style information learned by the model. We use the encoder decoder architecture to solve this problem of completing the set of fashion items. We evaluate the goodness of the proposed model through a variety of experiments. We empirically observe that our proposed model outperforms competitive baseline like apriori algorithm by \url{~28}\% in terms of accuracy for top-1 recommendation to complete the fashion ensemble. We also perform retrieval based experiments to understand the ability of the model to learn style and rank the complementary fashion items and find that using attention in our encoder decoder model helps in improving the mean reciprocal rank by \url{~24}\%. Qualitatively we find the complementary fashion items generated by our proposed model are richer than the apriori algorithm. 
\end{abstract}

%
%


\maketitle

\section{Introduction}
Fashion is a language that instantly conveys the persona. It is a choice with taste and styles, varying with space and time. It is a multi-billion dollar industry providing a surging market for e-commerce retailers, fashion designers and garment companies\footnote{https://www.statista.com/statistics/279757/apparel-market-size-projections-by-region/}. Like in many other domains, data driven technologies are making a difference in the fashion world~\cite{Lops2011,Aggarwal2016,McAuley_2015_C,He:2016:FFG:2872518.2890534,Jing:2015:VSP:2783258.2788621}, to name a few.


However fashion, along with its variations and personalizations, is not trivial to model. In particular the problem of modeling style is difficult in nature. Style depends on the ensemble of clothes worn, where the attribute of the clothes: color, pattern, demand attention. A pant may look stylish with a shirt, but a pink pant may not be considered stylish to wear with a yellow shirt. An abstract notion of style is formed when users prefer certain combination of apparels based on their attributes. This notion is often hard to evaluate, more so because of the subjectivity involved. Additionally, traditional shopping cart methods~\cite{Sarwar} do not work as well because there may be no relationship between clothes in the cart, e.g.,  pants could be bought to complement a blouse in the wardrobe, or for a family multiple items could be bought with no apparent relationships between purchases. Unlike books, movies and electronics, style keeps changing over time and becomes stale. Thus, purchase history for currently stylish items may not be available. 

\begin{figure}
		\centering
		\includegraphics[width=\columnwidth,height=2.8cm]{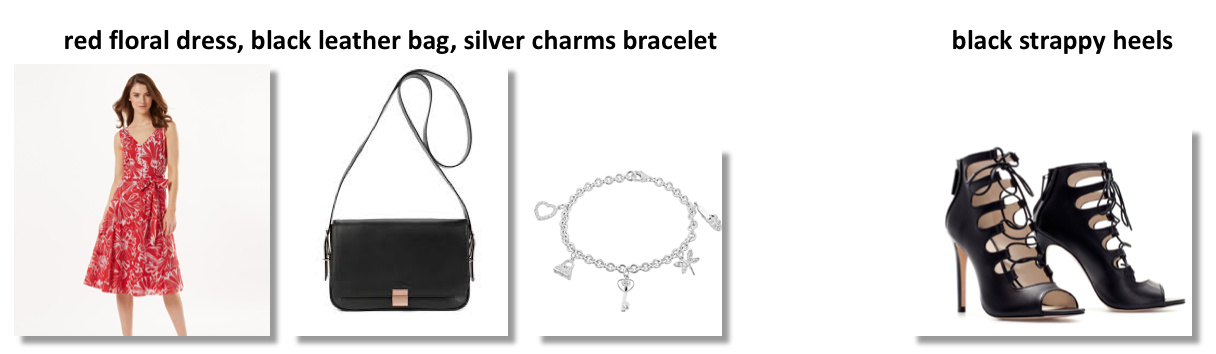}%
		\vspace{-6px}
		\captionof{figure}{Completing fashion ensembles (right) based on a  given set of fashion items.}
	\label{example_categories_1}%
\end{figure}

\begin{figure}
		\centering
		\includegraphics[width=\columnwidth]{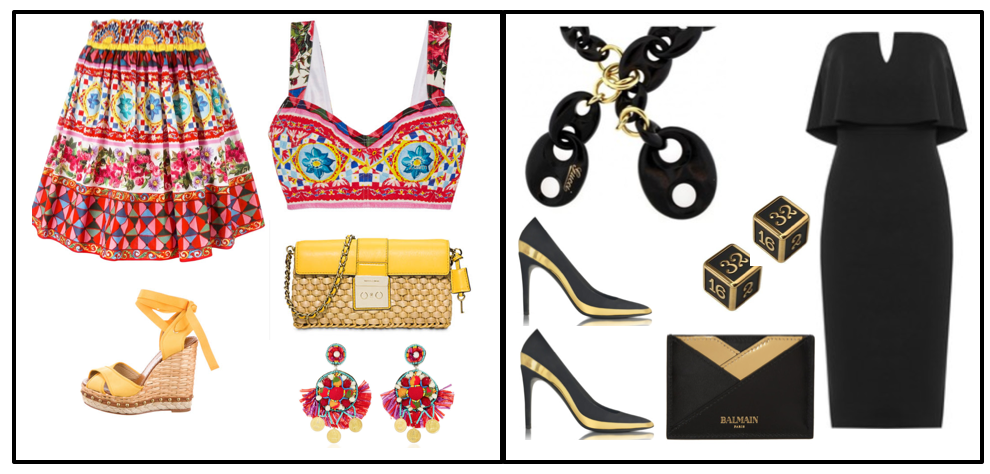}%
		\vspace{-6px}
		\captionof{figure}{Examples showing set of stylish fashion ensembles.}
	\label{example_ensemble}%
\end{figure}

\begin{figure*}
		\centering
		\includegraphics[width=\textwidth,height=3.5cm]{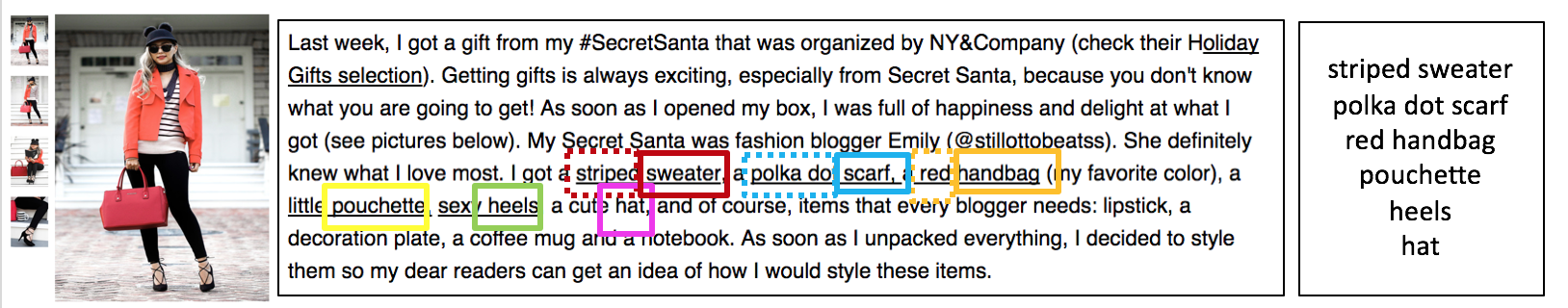}%
		\vspace{-6px}
		\captionof{figure}{Example post on social media by a user: the attributes (denoted by dashed lines) along with the apparels (denoted by solid lines) are marked in the same color. The box on the right hand side lists the set of items which appear in this post.}
	\label{socialpost}%
\end{figure*}

In this work we focus on the problem of completing fashion ensembles based on their style quotient. In a fashion ensemble there are complementary items that can be worn together, e.g.,  in Figure~\ref{example_categories_1},  for a given set of items \textit{red floral dress, black leather bag and silver bracelet}, an example complementary fashion items will be \textit{black strappy heels}. This is different from recommending similar items, that is given a red dress we are {\em not} recommending similar dresses.  Our problem on the other hand is given an ensemble of apparels, find stylish complementary items that could be worn with this ensemble. Note, completion here is used loosely -- it is not that the complement of our stylish set is not-stylish, or even that we cannot add more apparels to it; in fact our solution would be based on adding one apparel at a time to the ensemble. Thus it is for the user to decide when the look is completed. Further, we use apparels as a superset of clothes that includes accessories such as footwear, jewelery, head-wear etc.  

Style is characterized by complex inter-relationship among the fashion attributes that are not straight forward to capture. Figure~\ref{example_ensemble} shows examples of set of fashion items that could be worn together according to fashion experts on Polyvore\footnote{\label{polyvore}\url{http://www.polyvore.com/}}. Each of these ensembles of fashion items follow a certain style {\em rule}, e.g., black evening dress goes well with other black items, while the pink skirt will go better with a yellow sandal.  


The examples in Figure~\ref{example_ensemble} are of curated style rule, where a fashion experts have put together ensembles that would look good based on current style. Such curated advice is often expensive, obviously subjective and mostly out of date. To alleviate these problems and to quantify the abstract concept of style we propose to use social signals such as likes and comments by users on social media. Signals can easily be weighted to make current ones more relevant. Similarly, signals from friends or people with good style could be up-weighted.

There has been a rapid growth of online fashion social media websites like Polyvore\footnotemark[\getrefnumber{polyvore}] and Chictopia\footnote{\label{chictopia}\url{http://www.chictopia.com/}} where fashion enthusiasts share their \textit{outfit of the day}\footnote{\label{ootd}http://en.wikipedia.org/wiki/Outfit\_of\_the\_day}. Figure~\ref{socialpost} shows an example of one such post. The user posts an image along with text to describe the outfit of the day. These posts are a rich source of style information. 

We propose a data-driven solution to complete the fashion ensemble problem. Our data is parsed from posts appearing in social media websites discussed above. In this work we focus only on the textual description of the post. However, images can also be used as a source of modeling style. Posts are collated based on their social signals and then mined to create frequent itemsets where apparels that are often worn together, in a stylish look, occur together. Our problem then amounts to the problem of building conditional probability models that given a set of apparels can predict an apparel that is complementary to, and goes well with the set\footnote{as stated before we don't generate the ensemble in one shot but complete the ensemble one at a time. Given a set of fashion items $S$, the system recommends a complimentary fashion item $I$. Now, one can query the model again, with a new set of fashion item $S \cup I$ to complete the ensemble}. The input set can be of variable length (cardinality), not only in terms of the number of items in set, but also in terms of number of attributes attached to each item. The presence of attributes make the problem complicated, so the left example in Fig.~\ref{example_ensemble} shows that color does not predict stylishness, unlike the right example in the same figure. Similarly, the predicted fashion item can also be a variable length output depending on the number of attributes associated with the fashion item.



In order to accommodate the variable length input and output, we pose our problem as a sequence to sequence task. The input sequence is a selected subset of fashion items while the output sequence is a complementary fashion item. To address this sequence to sequence task, we build our predictive model using the encoder decoder recurrent neural network(RNN) architecture~\cite{cho2014learning}. RNNs are natural for variable length sequence modeling~\cite{SutskeverVL14}, and as we shall show that using {\em attention}~\cite{bahdanau2014neural} along with RNN allows us to take care of the details of the attributes. The encoder-decoder architecture consists of two recurrent neural networks (RNN) that act as an encoder and a decoder pair. The encoder maps a variable-length source sequence to a fixed-length vector, and the decoder maps the vector representation back to a variable-length target sequence. In our case, the input set of fashion items are represented in a compact form using the encoder. We use this compact representation to generate the output sequence of a complementary apparel via the decoder. We train our model by using all the parsed fashion posts. We hypothesize that the model will be able to capture the style information expressed in these posts. Additionally, our model incorporates an attention mechanism to explicitly learn the more interesting attributes in the fashion items such as color and pattern in order to improve the performance. We use the trained model as a outfit completion system by predicting the complementary fashion item for a given set of fashion items.

In order to evaluate our system we work using two novel datasets: one curated from social media website Chictopia and another obtained from an e-commerce website. We perform empirical study on both these datasets, showing our proposed model outperforms competitive baseline such as apriori algorithm on accuracy by \url{~28}\% in terms of accuracy for top-1 recommendation to complete the fashion ensemble and mean reciprocal rank by \url{~24}\% for the task of retrieval. Qualitatively, we also find that the complementary fashion items given by our proposed model are richer than the apriori algorithm. To the best of our knowledge, our work is the first to utilize the social media posts and address the problem of completing \textit{style based complementary} fashion items for a given set of fashion items.

We make the following main contributions in this paper:

\begin{itemize}
\item We propose the problem of completing a fashion ensemble such that items in the ensemble go well with each other. 
\item We propose an attention based recurrent neural network model that models the variable length input set of fashion items and generates a complementary apparel with varied number of attributes.
\item We compare the performance of the proposed sequence to sequence model with baselines such as apriori algorithm and find the neural network based model outperforms the pattern mining based method.
\end{itemize}

	The rest of the paper is organized as follows. Section~\ref{sec:problem} formally describes the task of \textbf{completing outfit ensembles} while Section~\ref{sec:model} discusses our proposed \textbf{encoder decoder RNN model}. Section~\ref{sec:exp} experimentally validates our proposed approach on two novel datasets, one curated from social media website Chictopia and another obtained from an Ecommerce website. We finally present the related work in Section~\ref{sec:related} and conclude the paper in Section~\ref{sec:conclusion}.
\section{Problem Formulation}
\label{sec:problem}

Let us begin by defining our dataset, $\mathcal{D}=\{(x^i,y^i)\}^{|\mathcal{D}|}_{i=1}$, where $x^i=[{x^i_1},\dots,{x^i_{s_i}}]$ correspond to the ${s}_i$ words\footnote{We will represent each query word as an indicator vector ${x_i} \in \{ 0,1\}^{V^S}$ where $|V^S|$ represents the size of the (query) word vocabulary.} in selected set of fashion items $i$ and $y^i=[{y^i_1},\dots,{y^i_{t_i}}]$ corresponds to the ${t}_i$ words\footnote{Similarly, each recommended item word is represented as an indicator vector ${y_i} \in \{ 0,1\}^{V^T}$ where $|V^T|$ represents the size of the (item) word vocabulary.} from the complementary fashion item $i$ predicted by the model to complete the input set of fashion items . Our task is generate a sequence of words $y^i$ for the complementary fashion item to complete the query of the selected set of fashion items $x^i$. 

Consider Figure~\ref{example_categories_1}, where we have a sequence of words ``red floral dress, black leather bag, silver charms bracelet'' and the goal is to predict a fashion item such as ``black strappy sandals'' that complements the query set of fashion items. 
\begin{figure*}
	\centering
		\includegraphics[width=\textwidth]{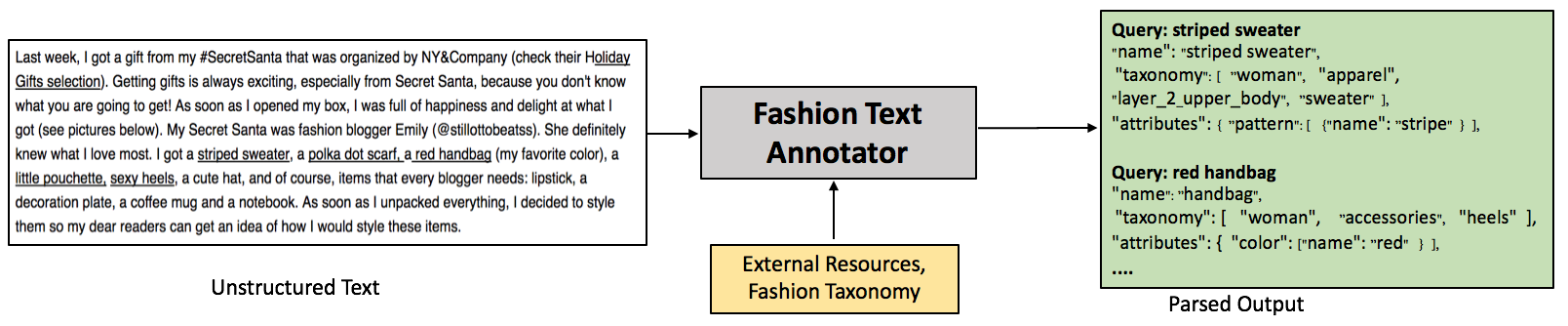}%
		\captionof{figure}{Working Principle of the Fashion Text Annotator}
	\label{fta}%
\end{figure*}

\begin{figure}
	\centering
		\includegraphics[width=\columnwidth]{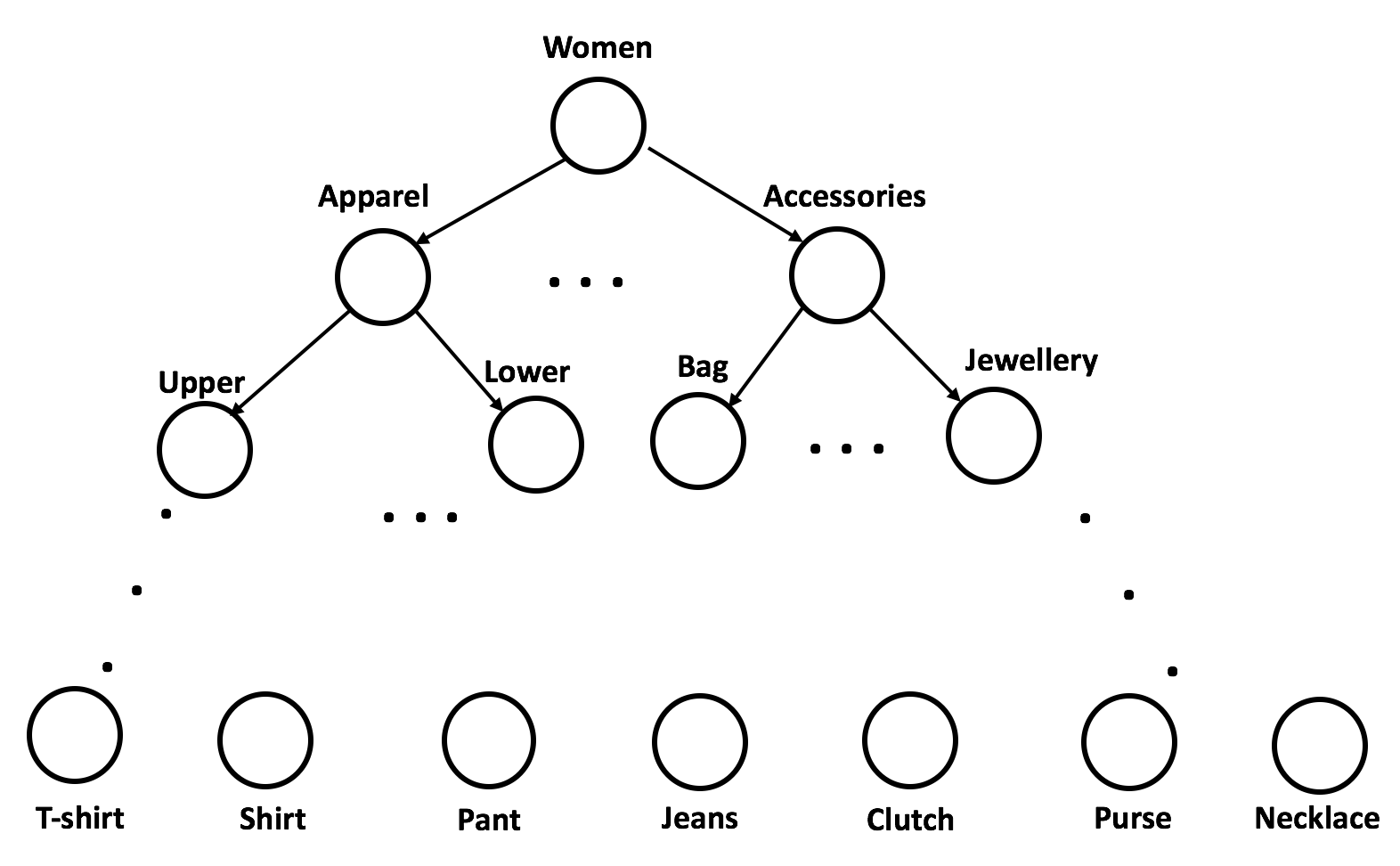}%
		\captionof{figure}{Fashion Taxonomy for Women}
	\label{taxonomy}%
\end{figure}

\begin{figure*}
		\centering
		\includegraphics[width=0.7\textwidth,height=7cm]{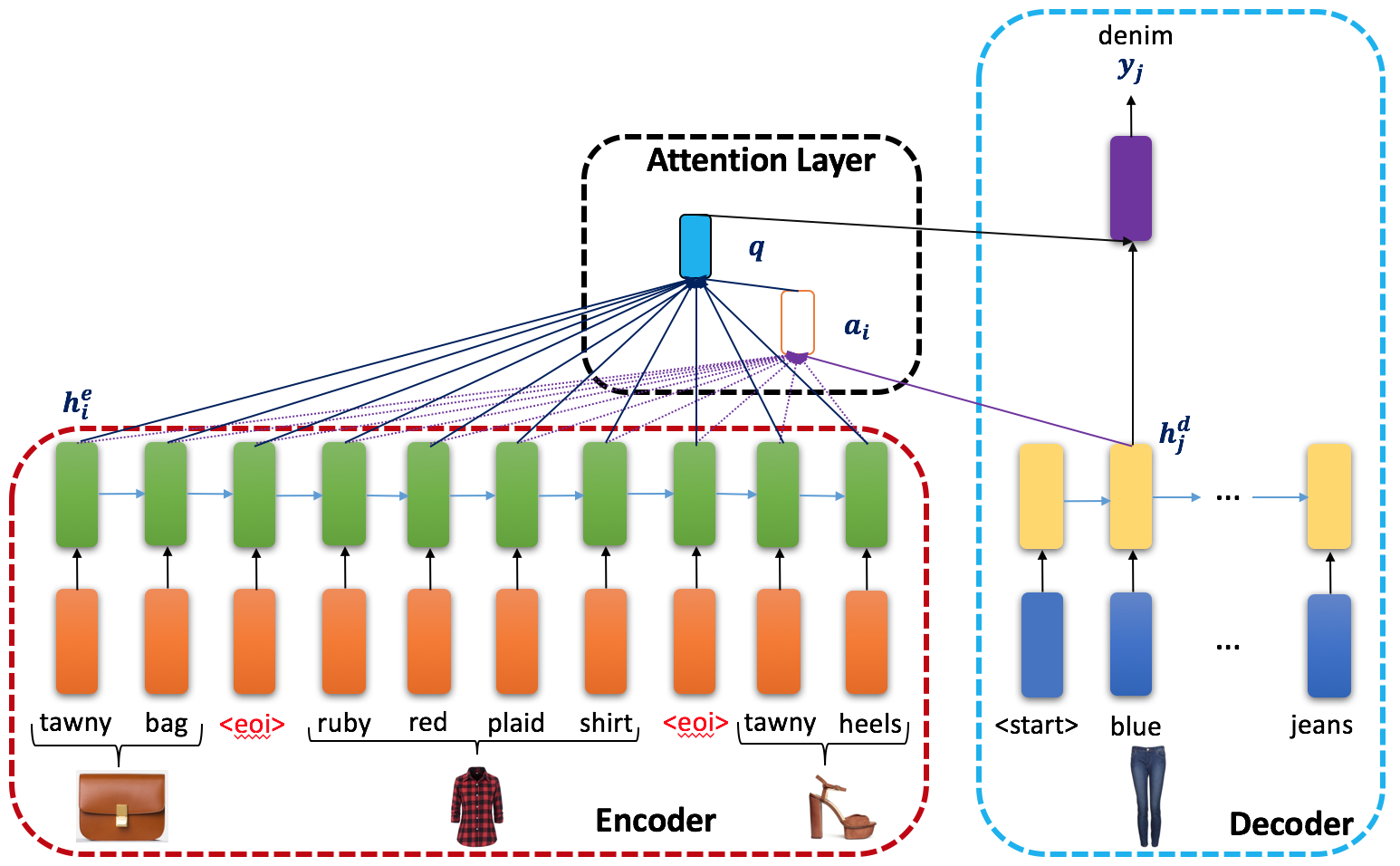}%
		\captionof{figure}{Proposed System: At each decoding time step j, the model infers a variable-length attention vector $a_i$ based on the current decoder state $h_d^j$ and all source states $h^e_i$. The input vector of a given set of fashion items q is then computed as the weighted average, according to $a_i$, over all the source states. The next word in the item is predicted based on the query vector q and current decoder state $h_d^j$. Note that each item in the encoder is separated by the `<eoi>' token to mark the item boundary. [Best viewed in color.]}
	\label{architecture}%
\end{figure*}

\section{Proposed Methodology}
\label{sec:model}
In this section we will discuss the proposed methodology in detail. Our approach can be broadly divided into two phases: 
\begin{itemize}
\item Mining style rules from social media
\item Modeling style using encoder decoder based RNN.
\end{itemize}

\subsection{\textbf{Mining style rules from social media}}
From social media websites, we crawl online fashion posts, $\mathcal{P}=\{p_{1},p_{2} \dots p_{n_p} \}$, where $n_p$ denotes the total number of posts. In order to filter the noise from these posts we use social signals such as likes and comments as a metric of the goodness of the fashion ensemble curated by these posts. We compute an average score of these social signals for every post and discard those posts which are below an empirically computed percentile. This, gives us a high quality set of fashion posts $\mathcal{S}=\{s_{1},s_{2} \dots s_{n_s} \}$, where $n_s < n_p$. 

We parse each of these \textit{unstructured} natural language post $s_{i}$ to obtain a \textit{structured} set of fashion items using a \textbf{Fashion Text Annotator (FTA)}. FTA uses a \textit{Fashion Taxonomy} (example shown in Figure~\ref{taxonomy}) and a set of external resources to parse unstructured natural language text into structured text. The FTA parses the unstructured post using an n-gram sliding window where $n$ ranges from 1 to 3. This value of n is chosen based on the average number of attributes per apparel in the dataset. For each n-gram it looks up the fashion taxonomy and the external list to extract the apparel along with its attributes such as color and pattern. Figure~\ref{fta} illustrates the working of the Fashion Text Annotator where the input is the unstructured user post mentioned in Figure~\ref{socialpost} and the output is structured data of apparels along with its attributes.

\subsection{\textbf{Modeling style using encoder decoder based RNN}}
Once we have the structured set of fashion items, we choose items from each of these structured set to generate the tuple $(x^i,y^i)$ where $y^i$ denotes the fashion item going well with the set of fashion items $x^i$. The goal is to learn a mapping of the style information (color, pattern, apparel type) encoded in these high quality fashion ensembles. Our proposed encoder decoder RNN model is illustrated in Figure~\ref{architecture} 

The model directly tries to maximize the conditional probability $p(y^i|x^i)$. Specifically, our model has two main components: (a) an \textit{encoder} which captures the essential information present in the input set of fashion items $x^i$ (b) a \textit{decoder} which outputs $y^i$, one target item word at a time of the complementary fashion item. If $D=|y^i|$, denotes the total number of tokens in the predicted complementary fashion item, the conditional probability can be decomposed as:
\vspace{-2px}
\begin{eqnarray}
\log p(y^i|x^i) = \sum^{|D|}_{j=1} \log p(y^i_j|y^i_{<j}, x^i)
\label{eq:cprob}
\end{eqnarray}

\noindent Our encoder and decoder will be discussed next in detail.

\textbf{\underline{Encoder:}}
The goal of the encoder (as shown in red dashed line in Figure~\ref{architecture}) is to represent the set of input items compactly. This compact representation is accessed by the decoder every time it emits a recommended item word. There are two main challenges in designing the encoder model. Firstly, the number of input symbols can be arbitrarily long in the query. This prohibits the usage of feedforward models which cannot model long-term dependencies~\cite{MikolovJCMR14} efficiently. In other words, it can only do so at the cost of a linear increase in the number of parameters with the increase in the number of input symbols considered. Secondly, the interactions of the different items in the input itemset along with their attributes such as color, pattern etc. can be very complex to be modeled thereby encouraging models with sophisticated architectures. The solution to the challenges will be discussed next.

Consider an itemset, $x=[{x_1},\dots, {x_{s}}]$ consisting of  ${s}$ words\footnote{\label{brevity} For the sake of brevity we ignore the superscript in this subsection}. For each query word $x_i$, our encoder RNN computes a dense vector called recurrent state, denoted by $h^e_i$, that combines $x_i$ with
the information that has already been processed so far, i.e. the
recurrent state $h^e_{i-1}$. Formally:
\vspace{-2px}
\begin{equation}
 \begin{aligned}
{h^e_i} = f({h^e_{i-1}}, {x_i})
\label{rnn}
\end{aligned}
\end{equation}

where $h^e_i \in R^{n}$, $n$ is the number of dimensions of the recurrent
state, $f$ is a non-linear transformation. The recurrent state $h^e_i$ acts as a compact summary of the words seen up to position $i$. Once Equation~\ref{rnn} has been run through the entire query $[{x_1},\dots, x_{s}]$, the last state $h^e_{s}$ may be viewed as a compact summary of the input query. 
We use the Long Short-Term Memory (LSTM) to reduce the fundamental difficulty in learning complex dependencies between fashion items and attributes, i.e. to store information for complex sequences.

The traditional encoder can consume a sequence of words present only in one item. To make our model practical, we propose a simple, yet effective compositionality technique for the encoder to consume multiple items. The idea is to concatenate all the words from different items separated by a special token (`$<$eoi$>$' marking the end of the item). Essentially, we combine all the item information to form one big input representing the itemset (user query) information. However, this simple strategy cannot be effective if the decoder is not able to give appropriate importance (or attention) to the input symbols consumed by the decoder.\\

\noindent \textbf{\underline{Decoder}}
 The goal of our decoder (as shown in blue dashed line in Figure~\ref{architecture}) is to predict a complementary fashion item word $y_j$ one at a time by accessing the encoder's top layer hidden states (which together captures the latent itemset embedding), eventually returning the item $y=\{y_1,\dots y_t \}$\footnotemark[\getrefnumber{chictopia}]. There are two main challenges in designing the decoder model. Firstly, the decoder must be equipped with an efficient mechanism to identify the area of focus per target word. The possible number of encoder positions can be very large (as we have merged multiple items from the itemset). Thus, the problem of locating the position of interest in the encoder corresponding to the salient information for the decoder with respect to the item to be predicted is non-trivial. For example, when the decoder is predicting the color attribute, it should emphasize on the position of the color words in the encoder. Secondly, the decoder has to sample from an extremely large set of candidate words.

Our decoder employs an LSTM model which parameterizes the probability of decoding each word $y_j$ as:
\vspace{-2px}
\begin{eqnarray}
p(y_j|y_{<j}, x) = softmax(W^d_{sf} h^d_j)
\label{eq:dec_1}
\end{eqnarray}
\vspace{-2px}
\noindent with $W^d_{sf}$ denoting the softmax weight matrix of size $|V^T| \times n$.

To utilize the information present in the input set of items spanning across several memories (green blocks in Figure~\ref{architecture}) in the encoder effectively, the decoder derives a query embedding $q$ using attention-based mechanism (which will be discussed next). This query embedding captures the salient information in the input that is useful to predict the current target word $y_j$. Precisely, we employ a simple concatenation layer to combine $q$ and $h^d_j$ to produce an attentional hidden state, which is then fed through the softmax layer to produce the predictive distribution formulated as:
\vspace{-2px}
\begin{eqnarray}
p(y_j|y_{<j}, x) = softmax(W^d_{sf}[q;h^d_j;])
\label{eq:dec_2}
\end{eqnarray}
\vspace{-2px}
To derive the query embedding ${q}$, we define a variable-length attention vector ${a_i}$, whose size equals the number of input word consumed by the encoder. We compare the current target hidden state $h^d_j$ with each encoder hidden state $h^e_i$ using dot product as:
\vspace{-2px}
\begin{eqnarray}
a_i = softmax({h^d_j}^\intercal h^e_i)
\label{eq:dec_3}
\end{eqnarray}
\vspace{-2px}
In Figure~\ref{architecture}, the block within black dashed line represents the attention layer of our model. Intuitively $a_i$ captures the degree to which a particular item word (encoder input symbol) helps to predict the next target word (say attribute value for color). For our fashion item set completion task, this gives a provision to negate the influence of irrelevant words by setting ${a_i}$ closer to 0 and encourage the influence of relevant words by setting the same closer to 1. Finally, the query embedding $q$ is computed as the weighted average over all the encoder hidden states, where the weights are given by $a_i$.

\noindent \underline{textbf{Model Optimization}}
Our model uses cross-entropy loss as the cost function and can be trained end-to-end by minimizing the negative conditional log likelihood (NLL) of the training data with respect to $\theta$:
\vspace{-2px}
\begin{eqnarray}
NLL(\theta) = \sum_{(x,y) \in \mathcal{D}}  - \log p(y|x,\theta)
\label{eq:obj}
\end{eqnarray}
\vspace{-2px}
Here, $\theta$ constitutes the encoder and decoder parameters. Once the model is trained we generate the complementary fashion item for a new set of fashion items $x$ through a word-based beam search such that $p(y|x)$ is maximized. The beam search reduces the impact of candidate explosion to a greater extent and is parameterized by the number of best paths $k$ that are pursued at each time step. We use Stochastic Gradient Descent (SGD) to learn the parameters of our model.

\section{Experimental Validation}
\label{sec:exp}
Evaluation of fashion based data driven system has always been a challenging task. For our problem of fashion item set completion,  the problem is further compounded with the introduction of style due to its subjective nature. In this section we discuss our datasets, baseline algorithm and experimental setup and finally present the quantitative and qualitative results. 

\begin{figure*}
\centering
\begin{tabular}{cc}
	\subfigure[ED]{\includegraphics[width=0.39\textwidth]{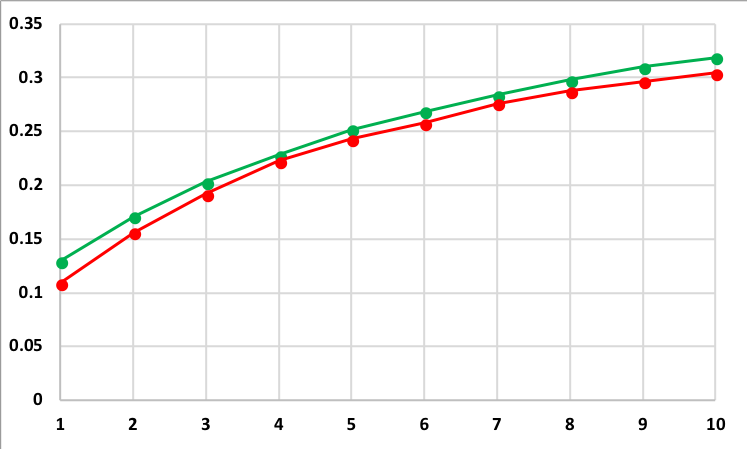}}
   	\subfigure[CD]{\includegraphics[width=0.39\textwidth]{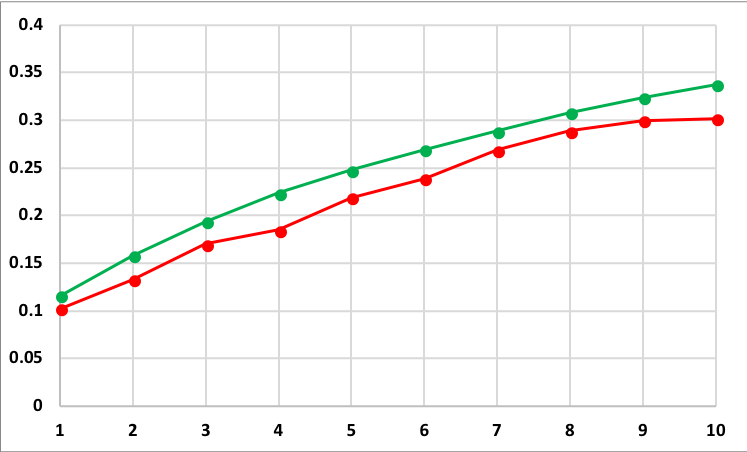}}
    \\
    \includegraphics[width=0.25\textwidth,height=0.45cm]{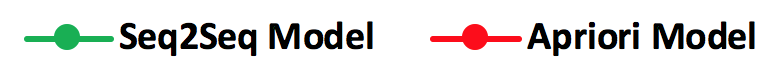}
\end{tabular}
\vspace{-10pt}
\caption{Jaccard Similarity Score: the x axis indicate the number of recommendation and y axis is the JSS score}
\label{js_all}
\end{figure*}
\vspace{-3px}
\subsection{Dataset}
Currently, there is no publicly available dataset which can be used for evaluating our solution against the problem of completing the set of fashion items. In order to evaluate the proposed solution we work with two real world datasets: social media posts from Chictopia\footnotemark[\getrefnumber{chictopia}] and style tips from an e-commerce website. \\

\noindent \textbf{\underline{Chictopia Dataset (CD):}} Chictopia is an online fashion portal which allows users to post their outfit of the day. Every post is associated with a free text description of the post along with various social activities like votes, comments and likes. We crawl about 0.15 million posts from Chictopia and parse these posts using our \textit{Fashion Text Annotator}. However, Chictopia like any other social media is noisy in nature. Many of these posts may be fashion blunders and therefore contain apparels which do not go well together. In order to filter these noisy posts, we exploit the wisdom of the crowd by defining a \textit{fashion score} for every post. The fashion score of a post is a weighted combination of number of votes, likes and comments. We take only those posts which are in the top 30 percentile based on this fashion score to filter the noisy posts. Finally, we obtain an automatically crowd-sourced \textit{~28K} golden fashion posts. The dataset consists of 135 unique colors, 95 unique patterns and 300 unique apparels.\\

\noindent \textbf{\underline{E-commerce Dataset (ED):}} We crawl the manually curated style tips curated by fashion designers for every item in the catalog from an e-commerce website. We parse these style tips using our Fashion Text Annotator to obtain \textit{10K} high quality style tips from the catalog. The dataset consists of 90 unique colors, 40 unique patterns and 238 unique apparels. The apparel along with the style tip would give us a set of attributed items going well together. Unlike  CD,  ED is noise free and does not require any filtering as it is manually curated by experts.\\

\noindent Note, for the sake of simplicity, we only consider posts and tips associated with women fashion and focus on the attributes color and pattern. Table~\ref{datastat} summarizes the number of train, validate and test posts for both the datasets.

\begin{table}
  \centering
  \caption{Statistics of Dataset}
  \label{datastat}
  \begin{tabular}{|c|c|c|c|c|}
  \hline
  	Dataset & Total & Train & Test & Validate \\ 
  	\hline
  	E-commerce Dataset & 10749 & 7524 & 2149 & 1076 \\
   	Chictopia Dataset & 27303 & 19112 & 5460  &  2731\\
  	\hline
    \end{tabular}
\end{table}

\subsection{Baseline Model}
Here we discuss the baseline algorithm which is used for comparing the performance of our encoder decoder RNN model. \\

\textbf{Apriori Algorithm:} The problem of finding items which go well together falls in the classical paradigm of frequent pattern mining. The \textit{apriori algorithm}~\cite{agrawal1993mining} is an influential algorithm to solve the problem of finding frequent patterns. We model the problem of generating stylish fashion ensembles as a frequent pattern mining problem. Given the dataset $\mathcal{D}$ (discussed in Section~\ref{sec:problem}), we employ apriori algorithm to mine itemsets to build a Style Rule Lexicon. Our lexicon consists of an attributed item and a list of attributed items which go well with it along with a support value. We further build this lexicon at different levels of granularity of attributes, i.e. considering all attributes, considering color or pattern only and considering no attribute at all. We use a minimum support value of 0.6 in case of CD to find frequent patterns while the support value for ED is 1 as the occurrence of every set of itemset is manually curated and validated by experts. 

\begin{figure*}
\centering
\begin{tabular}{cc}
	\subfigure[ED]
    {\includegraphics[width=0.39\textwidth]{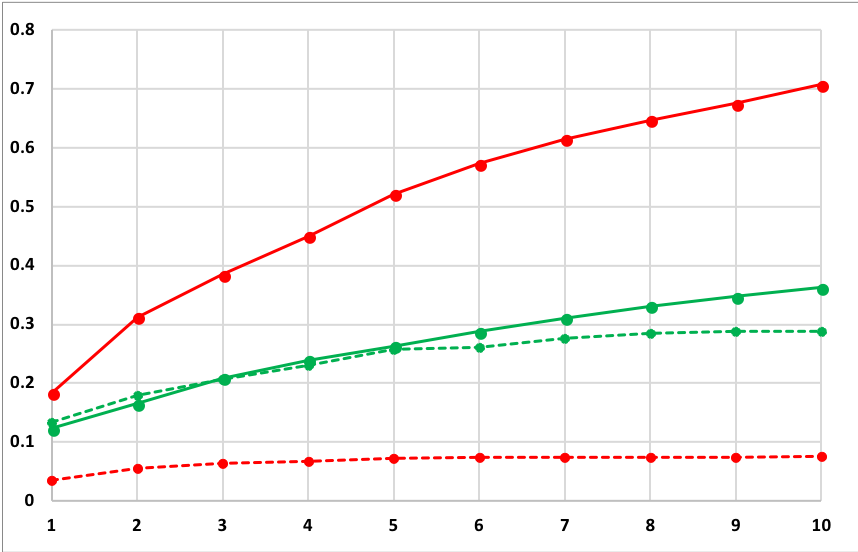}}
   	\subfigure[CD]{\includegraphics[width=0.39\textwidth]{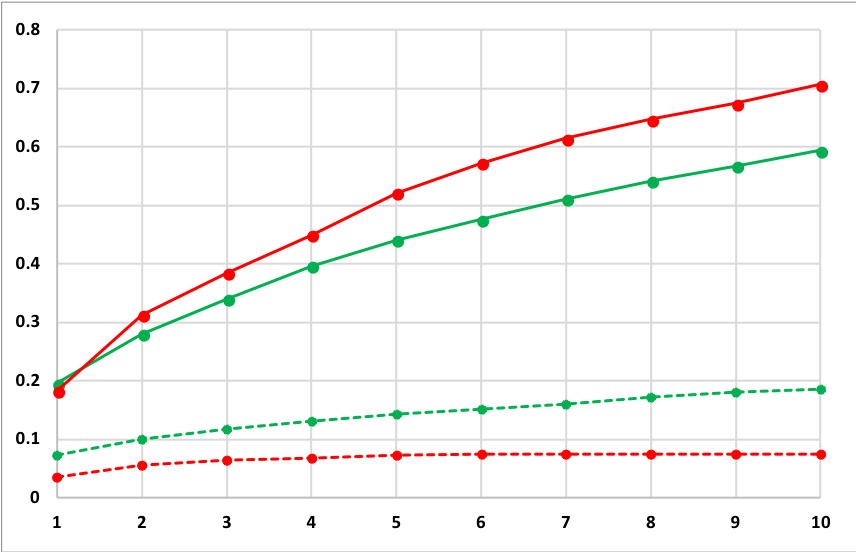}}
   \\[-2pt]
   \includegraphics[width=0.7\textwidth]{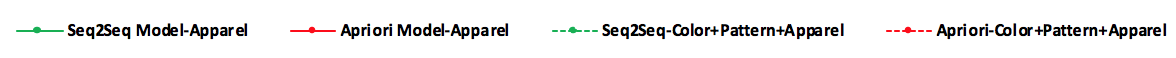}
\end{tabular}
\caption{Jaccard Similarity Score for CD: the x axis indicate the number of recommendation and y axis is the JSS score}
\vspace{-10pt}
\label{attributed_cd}
\end{figure*}

\subsection{Results}
In this section we evaluate our proposed algorithms (Seq2Seq Model) with the baseline (Apriori Model) for different quantitative tasks and present qualitative results. 

\subsubsection{Quantitative Results}\mbox{}\\
Quantitative evaluation in the field of fashion is difficult due to the subjective and abstract nature of it. We devise an experimental setup to perform three quantitative experiments to compare our algorithms against the baseline model for both the datasets. These are \textit{prediction accuracy}, \textit{generalisability} and \textit{retrievability}. We discuss each of these in the next section.\\

\underline{\textbf{Prediction Accuracy:}}
\label{attribute_original}
We use Jaccard Similarity Score (JSS) of the predicted attributed apparel (P) with the actual attributed apparel (A) for both the baseline model and our proposed model. We compute JSS for top-k predictions, where k ranges from 1 to 10. The  $JSS(A,P_i)@k$ for the predicted apparel $(P_i)$ and the ground-truth apparel $(A)$ where $1 \leq i \leq k$ is given as follows:

\begin{equation}
	JSS(A,P_i)@k = \max_{1<=i<=k}\frac{A \cap P_i}{A \cup P_i}
\end{equation}

Figure~\ref{js_all} illustrates the performance of the different models for both ED and CD respectively. We observe that Seq2Seq model performs consistently better than Apriori model across both the datasets giving an improvement of \url{~40}\% and \url{16}\% for CD and ED respectively. This shows that our model benefits from attending to important attributes in the input set of fashion items. We can further see that the gain for sequence to sequence model with respect to apriori is more in case of CD. We see that the ED dataset has less number of training data and fairly large number of fashion attributes. Therefore our model is not able to learn style from ED data as well as it does for CD data.\\

In order to understand the complexity of the attributes while generating style rules we compute JSS@k at different levels of granularity, by taking different set of attributes at a time. We consider color and pattern as the attributes and exploit the following combinations: color+pattern+apparel and apparel only. \\

Figures~\ref{attributed_cd} illustrate the performance of the system for different level of granularity for both the datasets. We find that recommending attribute based item is more challenging. The apriori mining algorithm fails in generating color+pattern+apparel recommendation while Seq2Seq Model is able to beat the Apriori model for both the datasets. A similar trend is observed for color+apparel and pattern+apparel task where our proposed model retains its supremacy. (We don't include the graphs due to limited space constraints). This performance is consistently observed for both the datasets. We observe that Apriori model performs well when compared with our proposed model for the task of apparel prediction. This is because of limited examples of apparel only examples seen by  our sequence to sequence model. While training our sequence to sequence model, the model is not able to see enough examples to learn. Nevertheless, the performance of the apriori algorithm starts dropping as we add more attributes and our proposed model is the winning model.\\\\

\begin{figure*}
\centering
\begin{tabular}{cc}
	\subfigure[Performance on CD when model is trained on ED]{\includegraphics[width=0.39\textwidth]{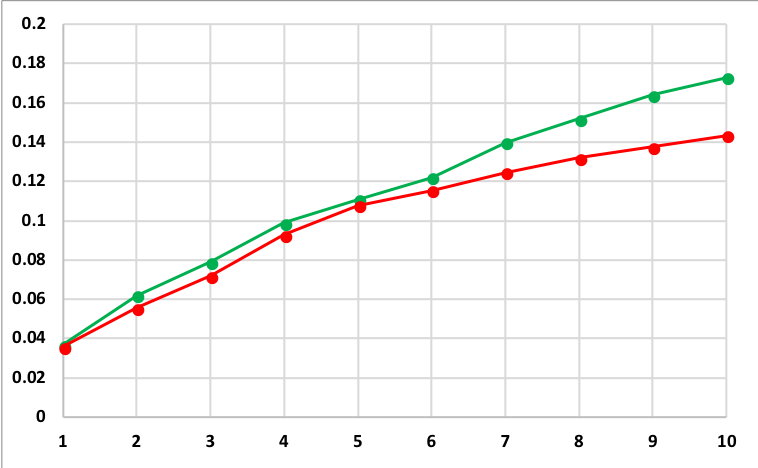}}
   	\subfigure[Performance on ED when model is trained on CD]{\includegraphics[width=0.4\textwidth]{./results_v2/CD_to_ED_all.png}} 
 \\[-2pt]
 \includegraphics[width=0.3\textwidth]{./results_v2/legend_all.png}
\end{tabular}
\caption{Jaccard Similarity Score for Evaluating Generalization: the x axis indicate the number of recommendation and y axis is the JSS score }
\vspace{-10pt}
\label{js_cross}
\end{figure*}

\begin{figure*}
\centering
\begin{tabular}{cc}
	\subfigure[ED]{\includegraphics[width=0.4\textwidth]{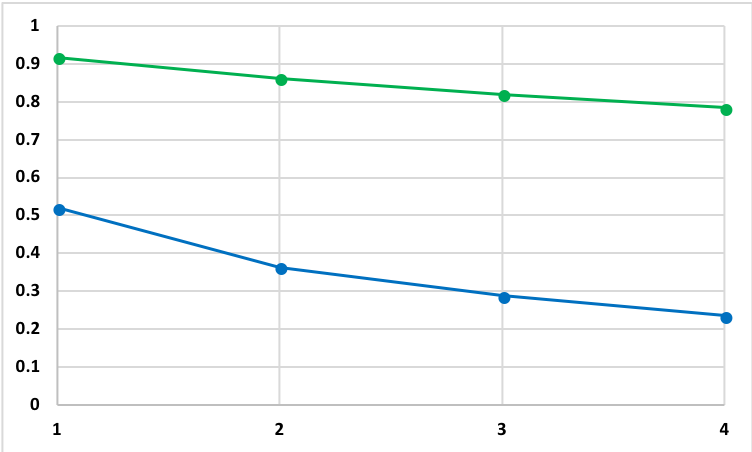}}
   	\subfigure[CD]{\includegraphics[width=0.4\textwidth]{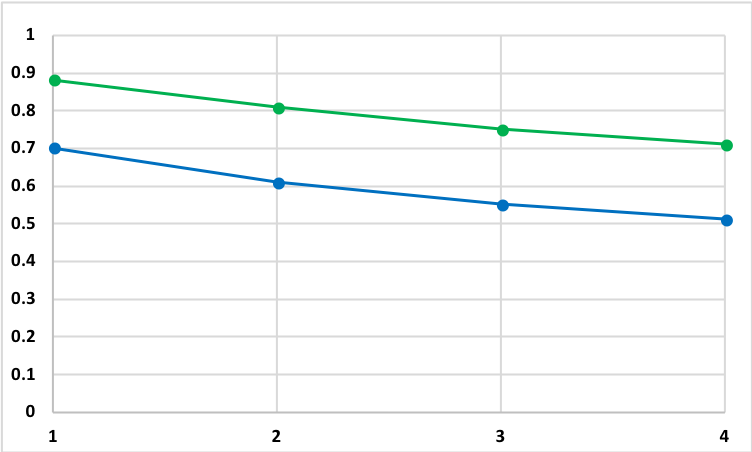}}
       \\[-2pt]
   \includegraphics[width=0.45\textwidth]{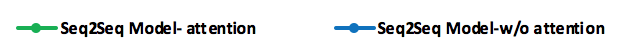}
\end{tabular}
\caption{Mean Reciprocal Rank(MRR): The x axis indicate the number of negative samples taken and the y axis indicate the MRR}
\vspace{-10pt}
\label{ed_mrr}
\end{figure*}

\underline{\textbf{Generalization:}}

We study the generalization capability of the proposed models. In this experiment we examine the performance of the model by taking data from a different distribution but in the same domain. In order to do this, we use CD test data to evaluate the model trained on ED and vice-versa. This enables us to measure how well the model can transfer knowledge to newer test data when the underlying distribution changes. We compute JSS@k for different algorithms for both the datasets. Figure~\ref{js_cross} indicates the JSS@k score on CD using model trained on ED and the JSS@k score on ED using model trained on CD. We find that in this transfer learning evaluation task, apriori algorithm performs well when using model trained on ED and tested on CD. This is because our model is not able to learn well on the limited data of ED which does not have enough examples of the itemsets repeating in the training data. The problem is alleviated in case of apriori algorithm since in case of apriori algorithm we use the minimum support as 1. However when we train using CD, which is richer in terms of attributes and data points, we find that our proposed model beats the apriori model. \\

\underline{\textbf{Retrieval Based Experiments:}}

While JSS@k is a good indicator to compare the accuracy of the model for the task of prediction, we evaluate the model on the task of ranking recommendations. The intuition behind this experiment is to evaluate the capability of the model to rank the correct fashion item higher than an arbitrary recommendation. If the model is capable to perform well in this task, one can infer that the performance of the model is not random and it is indeed learning from the data.\\

Consider a query apparel $(x,l)$ where $l$ is the corresponding \textit{ground-truth label}. We perform uniform random sampling to obtain $k$ negative samples $\{y^1, \dots y^k\}$ from the set of all ground-truth labels. For a given model, we compute the conditional probabilities $P(l|x)$  and $\forall i \in \{1,k\}, P(y^i|x)$ to predict the labels $l$ and $\{y_i\}$ respectively, for the same query $x$. We rank the ground-truth label $l$ and randomly chosen set of $k$ recommendations, $\{y^{i}\}$ in descending order of their conditional probability scores. In an ideal case, the rank $r_{l}$ of the ground-truth label $l$ should be more than the rank $r_{y^i}$ of the arbitrary negative samples. We compute the performance of ranking through the statistic measure Mean Reciprocal Rank (MRR) given by Equation~\ref{mrr}. MRR gives a measure of the predicted rank for the true label, which in an ideal case should be 1. 

For the task of retrieval based experiment we cannot use apriori algorithm due to its deterministic nature. Therefore to compare our proposed model we use a variant of our model called the \textit{Seq2Seq model without Attention}. This model does not use the attention mechanism during training the model.

\begin{equation}
\label{mrr}
 MRR = \frac {1}{|n|} \sum _{{i=1}}^{{|n|}}\frac  {1}{rank_i}
\end{equation}

Figure \ref{ed_mrr} illustrates the MRR for both the datasets for the Seq2Seq with attention and Seq2Seq without attention based models. We vary $k$, the number of negative samples from 1 to 4. Note that the probability given by a randomly performing model is $\{0.5,0.33,0.25,0.2 \}$ for $k = \{1,2,3,4\}$ respectively. We find that the Seq2Seq Model is able to rank the correct complementary fashion item at rank 1 with high value of MRR for both the datasets. We also observe that the Seq2Seq with attention model outperforms Seq2Seq without attention model resulting in an average improvement of $~19\%$ and $~36\%$ respectively. This experiment demonstrates the importance of attention while decoding the attributes of the fashion items. 

\subsubsection{Qualitative Results}\mbox{}\\
In order to validate the quality of the generated complementary fashion items we present the qualitative results for our proposed model and the baseline method. Tables~\ref{styletip} and ~\ref{styletip1} illustrates some example queries and the predicted complementary fashion item for different models on ED and CD dataset respectively. Note that for many esoteric attributes values like ivory color and in case of heavily attributed set of fashion items, the Apriori Model fails to generate any complementary fashion item. In contrast our proposed Seq2Seq model is able to generate good quality fashion items to add in the query set of fashion items.

\begin{table*}{}
  \centering
  \caption{Qualitative Results for Fashion Item Prediction for ED Dataset}
  \label{styletip}
  \begin{tabular}{|p{4.5cm}|p{4cm}|p{4cm}|p{4cm}|}
  \hline
\textbf{Input Set of Fashion Items}  & \textbf{Apriori Algorithm } &   \textbf{Seq2Seq Model w/o Attention}  & \textbf{Seq2Seq Model with Attention} \\ \hline           
blue printed jeans
 &  NIL & black t-shirt & black solid top \\ \hline
medium stone blue printed kurta, brown clutch &  red printed skirt & sandals & copper toned sandals \\ \hline
white crop top, grey joggers &  white sneakers & running shoes & black running shoes \\ \hline
maroon camisole top
 &  NIL & casual shoes & white trousers \\ \hline
blue printed leggings, white heels &  NIL & black printed kurta & white printed kurta \\ \hline

\end{tabular}
\end{table*}

\begin{table*}{}
  \centering
  \caption{Qualitative Results for Fashion Item Prediction for CD Dataset}
  \label{styletip1}
  \begin{tabular}{|p{3cm}|p{4.5cm}|p{4.5cm}|p{4.5cm}|}
  \hline
\textbf{Query}  & \textbf{Apriori Algorithm } &   \textbf{Seq2Seq Model w/o Attention}  & \textbf{Seq2Seq Model with Attention} \\ \hline           
black polka dot tights &  NIL & black dress & black lace dress \\ \hline
yellow print jacket, brown leather boots &  NIL & dress & black skirt \\ \hline
black tights, mustard cardigan, brown boots, white blouse & gloves & black short & black printed skirt \\ \hline
white woven shirt, light blue trousers &  black pumps & blue bag & ivory printed coat  \\ \hline
navy trench coat &  NIL & white top & blue dress \\ \hline
\end{tabular}
\end{table*}

\section{Related Work}
\label{sec:related}

This work is closely related to two sub-fields: application of recommender systems and deep learning.\\

\noindent \textbf{\underline{Recommender System:}} The closest area of research is complementary item recommendation which has gained a lot of interest from the researchers in the field of recommender systems~\cite{McAuley_2015_C,Kalantidis:2013,McAuley_2015,Veit_2015_ICCV}. Veit et. al~\cite{Veit_2015_ICCV} propose single item recommendation for a given item. They learn a feature transformation from images of items into a latent space that expresses compatibility and model pairwise compatibility based on co-occurrence in large-scale user behavior data; in particular co-purchase data from Amazon.com. McAuley et. al~\cite{McAuley_2015} proposes joint recommendation of complimentary and substitutable products by formulating it as a supervised link prediction task. They employ product reviews to build topic models to learn such relationships. McAuley et. al~\cite{McAuley_2015_C} proposes image based recommendations for recommending clothes and accessories that go well together based on visual cues. In ~\cite{Kalantidis:2013} the authors propose an approach to learn relationships between clothing items and events (e.g. birthday parties, funerals) in order to recommend event-appropriate items. They learn a supervised model using visual features on a predefined set of categories and attributes. Although related to our problem, these methods require handcrafted methods and carefully annotated data for recommending clothing categories for a given occasion. 

Fashion is driven by style and is set by fashion enthusiasts evolving over time. Most of the above discussed recommender systems are based on using the product images, purchase history or reviews of the product. However, these aspects do not capture the fashion aesthetics and are therefore incapable to recommend complementary stylish product.  To the contrary, we exploit the goldmine of social media from fashion experts thereby learning the aesthetics of fashion while leveraging the fashion taxonomy to understand style which forms the backbone of fashion.  \\

\noindent \textbf{\underline{Deep Learning:}} Deep learning has excelled in providing the state-of-the-art models in diverse applications such as machine reading and comprehension~\cite{NIPS2015_5945}, machine translation~\cite{45610}, query suggestion~\cite{Sordoni:2015} and summarization~\cite{Abigail}. This technology helps in building models with multiple desirable features: (a) minimal feature engineering (b) ability to create expressive models and (c) minimal assumptions about the domain thereby enabling easier portability to newer domains~\cite{Bengio:2013}. This inspires us to tap its potential for building accurate models for capturing style and trend. For our complementary item recommendation problem, we have used the sequence-to-sequence model~\cite{SutskeverVL14} which has shown significant improvements in word error rates for conditional text generation problems such as machine translation~\cite{45610} and long text summarization~\cite{Abigail}. Our expressive recommendation model provides predictions which are not only accurate but are generalizable. We believe our work would revive the interest among the researchers to apply deep learning to solve challenging problems in the fashion domain.

\section{Conclusion and Future Work}
\label{sec:conclusion}

In this work, we formally defined the problem of completing set of fashion items and proposed a sequence to sequence algorithm to solve this task. Finally, we applied this algorithm to the hitherto task of generating stylish fashion ensembles and demonstrated the efficiency of the system both quantitatively and qualitatively. In future, we would like to explore other information accompanying the post like comments from users sentiment in these comments, images etc. thereby improving the quality of recommendation. We would also like to investigate a different loss function for the proposed model addressing the subjective nature of the task. Finally, we want to move from  predicting an item to an \textit{itemset} prediction that can enumerate all the recommended items in one shot. The challenge of predicting an itemset is difficult to be optimized using cross entropy loss and it would be interesting to explore reinforcement learning to tackle this bottleneck.

\bibliographystyle{ACM-Reference-Format}


\begin{thebibliography}{0}


\ifx \showCODEN    \undefined \def \showCODEN     #1{\unskip}     \fi
\ifx \showDOI      \undefined \def \showDOI       #1{#1}\fi
\ifx \showISBNx    \undefined \def \showISBNx     #1{\unskip}     \fi
\ifx \showISBNxiii \undefined \def \showISBNxiii  #1{\unskip}     \fi
\ifx \showISSN     \undefined \def \showISSN      #1{\unskip}     \fi
\ifx \showLCCN     \undefined \def \showLCCN      #1{\unskip}     \fi
\ifx \shownote     \undefined \def \shownote      #1{#1}          \fi
\ifx \showarticletitle \undefined \def \showarticletitle #1{#1}   \fi
\ifx \showURL      \undefined \def \showURL       {\relax}        \fi
\providecommand\bibfield[2]{#2}
\providecommand\bibinfo[2]{#2}
\providecommand\natexlab[1]{#1}
\providecommand\showeprint[2][]{arXiv:#2}

\end{thebibliography}


\begin{thebibliography}{10}

\bibitem{Lops2011}
Pasquale Lops, Marco de Gemmis, Giovanni Semeraro
\newblock 2011.
\newblock Content-based Recommender Systems: State of the Art and Trends
\newblock {\em Recommender Systems Handbook}, 73--105.

\bibitem{Aggarwal2016}
Charu C. Aggarwal
\newblock 2016.
\newblock Content-Based Recommender Systems
\newblock {\em Recommender Systems: The Textbook}, 139--166.

\bibitem{McAuley_2015_C}
Julian McAuley, Christopher Targett, Qinfeng Shi, Anton van den Hengel
\newblock 2015.
\newblock Image-Based Recommendations on Styles and Substitutes
\newblock {\em SIGIR}, 43--52	

\bibitem{He:2016:FFG:2872518.2890534}
Ruining He, Chunbin Lin, Julian McAuley
\newblock 2016.
\newblock Fashionista: A Fashion-aware Graphical System for Exploring Visually Similar Items
\newblock {\em WWW Companion}, 99--202.

\bibitem{Jing:2015:VSP:2783258.2788621}
Yushi Jing, David Liu, Dmitry Kislyuk, Andrew Zhai, Jiajing Xu, Jeff Donahue, Sarah Tavel
\newblock 2015.
\newblock Visual Search at Pinterest
\newblock {\em KDD}, 1889--1898

\bibitem{Sarwar}
B. Sarwar, G. Karypis, J. Konstan, J. Riedl
\newblock 2001.
\newblock Item-based Collaborative Filtering Recommendation Algorithms
\newblock {\em WWW}, 285--295.

\bibitem{cho2014learning}
Kyunghyun Cho, Bart van Merrienboer, Caglar Gulcehre, Dzmitry Bahdanau, Fethi Bougares, Holger Schwenk, Yoshua Bengio
\newblock 2014.
\newblock Learning phrase representations using RNN encoder-decoder for statistical machine translation
\newblock {\em EMNLP}

\bibitem{SutskeverVL14}
Ilya Sutskever, Oriol Vinyals, Quoc Le
\newblock 2014.
\newblock Sequence to Sequence Learning with Neural Networks
\newblock {\em NIPS}, 3104--31120

\bibitem{bahdanau2014neural}
Dzmitry Bahdanau, Kyunghyun Cho, Yoshua Bengio
\newblock 2015.
\newblock Neural machine translation by jointly learning to align and translate
\newblock {\em ICLR}


\bibitem{MikolovJCMR14}
Tomas Mikolov, Armand Joulin, Sumit Chopra, Micha{\"{e}}l Mathieu, Marc'Aurelio Ranzato
\newblock 2014.
\newblock Learning Longer Memory in Recurrent Neural Networks
\newblock {\em JCMR}

\bibitem{agrawal1993mining}
Rakesh Agrawal, Tomasz Imieli{\'n}ski, Arun Swami
\newblock 1993.
\newblock Mining association rules between sets of items in large databases
\newblock {\em SIGMOD}, 207--216

\bibitem{Kalantidis:2013}
Kalantidis, Yannis and Kennedy, Lyndon and Li, Li-Jia
\newblock 2013.
\newblock Getting the Look: Clothing Recognition and Segmentation for Automatic Product Suggestions in Everyday Photos
\newblock {\em  ICMR}, 105--112


\bibitem{McAuley_2015}
Julian McAuley, Rahul Pandey, Jure Leskovec 
\newblock 2015.
\newblock Inferring Networks of Substitutable and Complementary Products
\newblock {\em KDD}, 785--794	

\bibitem{Veit_2015_ICCV}
Andreas Veit, Balazs Kovacs, Sean Bell, Julian McAuley, Kavita Bala, Serge Belongie
\newblock 2015.
\newblock Learning Visual Clothing Style With Heterogeneous Dyadic Co-Occurrences
\newblock {\em ICCV}, 4642-4650

\bibitem{NIPS2015_5945}
Karl Moritz Hermann, Tomas Kocisky, Edward Grefenstette, Lasse Espeholt, Will Kay, Mustafa Suleyman, Phil  Blunsom
\newblock 2015.
\newblock Teaching Machines to Read and Comprehend
\newblock {\em NIPS}, 1693--1701

\bibitem{45610}
Yonghui Wu et al. 
\newblock 2016.
\newblock Google's Neural Machine Translation System: Bridging the Gap between Human and Machine Translation
\newblock {\em TACL}

\bibitem{Sordoni:2015}
Alessandro Sordoni, Yoshua  Bengio, Hossein Vahabi, Christina Lioma, Jakob Grue Simonsen, Jian-Yun Nie,
\newblock 2015.
\newblock A Hierarchical Recurrent Encoder-Decoder for Generative Context-Aware Query Suggestion
\newblock {\em CIKM}, 553--562

\bibitem{Abigail}
Abigail See, Peter J. Liu, Christopher D. Manning 
\newblock 2017.
\newblock Get To The Point: Summarization with Pointer-Generator Networks
\newblock {\em ACL}, 1073--1083

\bibitem{Bengio:2013}
Yoshua Bengio, Aaron Courville, Pascal Vincent, 
\newblock 2013.
\newblock Representation Learning: A Review and New Perspectives
\newblock {\em IEEE Trans. Pattern Anal. Mach. Intell.}, 1798--1828

\end{thebibliography}

\end{document}